\def\halpha{{\rm H}$\alpha$~}
\def\hbeta{{\rm H}$\beta$~}
\def\Ha{{\rm H}\alpha}
\def\Hb{{\rm H}\beta}
\def\etal{{\em et al.~}}
\def\bcr{\vspace*{0em}}

\def\ROIII{R_{\rm O \, III}}
\def\ROII{R_{\rm O \, II}}
\def\RSII{R_{\rm S \, II}}

\def\figdir{ }

\documentstyle[11pt,aaspp]{article}

\eqsecnum

\slugcomment{To appear in the {\em The Astrophysical Journal} (1997 Sept 1)}

\begin{document}

\title{A Deep Look at the Emission-Line Nebula in Abell 2597}

\author{G. Mark Voit and Megan Donahue}
\affil{STScI, 3700 San Martin Drive, Baltimore,
MD 21218}

\setcounter{footnote}{0}

\begin{abstract}
The close correlation between cooling flows and emission-line nebulae 
in clusters of galaxies has been recognized for over a decade and a
half, but the physical reason for this connection remains unclear.
Here we present deep optical spectra of the nebula in Abell~2597, one of 
the nearest strong cooling-flow clusters.  These spectra
reveal the density, temperature, and metal abundances of the 
line-emitting gas.  The abundances are roughly half-solar, and dust
produces an extinction of at least a magnitude in $V$.
The absence of [O III] $\lambda$4363 emission rules out shocks as
a major ionizing mechanism, and the weakness of He~II $\lambda$4686
rules out a hard ionizing source, such as an AGN or cooling
intracluster gas.  Hot stars are
therefore the best candidate for producing the ionization.
However, even the hottest O-stars cannot power a nebula as hot as
the one we see.  Some other non-ionizing source of heat appears
to contribute a comparable amount of power.  We show that the
energy flux from a confining medium can become important when 
the ionization level of a nebula drops to the low levels seen 
in cooling-flow nebulae.  We suggest that this kind of phenomenon,
in which energy fluxes from the surrounding medium augment 
photoelectric heating, might be the common feature underlying
the diverse group of objects classified as LINERS.
\end{abstract}


\section{Introduction}

A magnificent ionization nebula fills the central 20~kpc of
Abell~2597, a relatively nearby cluster of galaxies at
$z=0.0821$.  The H$\alpha$+[N~II] luminosity of this nebula,
uncorrected for reddening, is $2.7 \times 10^{42} \, 
{\rm erg \, s}^{-1}$ (Heckman \etal 1989). This 
intracluster emission-line display is not unique. 
Such nebulae are common among clusters whose 
central cooling times are shorter than a Hubble time 
but are absent in clusters with longer central cooling 
times (Hu, Cowie, \& Wang 1985; Heckman et al. 1989).
The strong correlation between central cooling time 
and optical nebulosity suggests that the cooling of the intracluster
medium somehow excites the line emission.  Yet, despite
this tantalizing hint, the true source of the ionizing 
energy in these clusters has remained mysterious.
 
The puzzle is not new.  Over 65 years ago Hubble \& Humason (1931)
had already noted the unusual appearance of NGC~1275, the central 
galaxy of the Perseus cluster.  It is quite blue for
an elliptical galaxy, and its spectrum bristles with strong
emission lines.  Two decades later, Baade \& Minkowski (1954) 
identified NGC~1275 with the radio source Perseus~A, and 
Minkowski (1957) discovered that its extended emission-line 
system separated into two subsystems differing by 3000~km~s$^{-1}$ 
in radial velocity.  Minkowski suggested that the two velocity 
subsystems arose from a galaxy-galaxy collision.  Soon thereafter,
the Burbidges speculated that NGC~1275 might be an exploding 
galaxy (Burbidge, Burbidge, \& Sandage 1963; Burbidge \& Burbidge 
1965), an interpretation given impetus by the the spectacular 
H$\alpha$ image of Lynds (1970) showing a complex of filaments 
extending up to 100~kpc from the galaxy's center.

X-ray astronomers came upon the mystery of emission lines
in clusters from a different direction.  Once X-ray telescopes
began to resolve the hot intracluster medium (ICM) in the 
nearest clusters, it became apparent that the central 
cooling times in clusters were frequently shorter than a Hubble time.
The brevity of these cooling times meant that, if there were no 
heat source to replenish the radiative losses, the ICM at 
the cores of such clusters must be cooling and settling 
towards the center in a ``cooling flow''   
(see Fabian 1994 for a review).  
Significantly, many of the earliest known 
cooling-flow clusters also displayed emission-line systems, 
and advocates of the cooling-flow hypothesis suggested 
that condensing intracluster gas somehow emitted the 
optical lines as it cooled from several times $10^7$~K down
through $10^4$~K (Fabian \& Nulsen 1977; 
Mathews \& Bregman 1978).

Quantitative followup of cooling-flow clusters revealed
that the connection between ICM cooling and line emission
was complicated. The H$\alpha$ luminosity of
a hot homogenous gas cooling through H$\alpha$-emitting 
temperatures is $\sim (3.8 \times 10^{39} \, {\rm erg \,
s^{-1}}) \dot{M}_{100}$, where 
$\dot{M}_{100}$ is the mass cooling rate 
in units of 100 solar masses per year.  
The H$\alpha$ luminosities of intracluster nebulae, if they
arose from simple cooling, would require $\dot{M}$ rates
up to $10^4 \, M_\odot \, {\rm yr^{-1}}$, tens to hundreds 
of times higher than the X-ray derived cooling rates. 
 
Even the more modest cooling rates inferred from X-ray observations, 
typically $\dot{M}_X \sim 10-1000 \, M_\odot \, {\rm yr^{-1}}$, face a 
crisis which continues: we have not yet figured out where 
the cold gas goes. Normal star formation in cooling-flow 
clusters progresses at only 1-10\% of $\dot{M}_X $
(e.g. McNamara \& O'Connell 1989, 1992; O'Connell \& McNamara 1989). 
Enigmatic soft X-ray absorption in many clusters seems to indicate 
a large mass of cold intracluster gas 
(White \etal 1991; Allen \etal 1993; Ferland, Fabian, \& Johnstone
1994), but this presumably molecular gas has not yet been
detected in any other waveband (McNamara, Bregman, \& O'Connell 1990;
O'Dea \etal 1994; Braine \etal 1995; 
Voit \& Donahue 1995).

While these difficulties with the cooling-flow hypothesis leave
ample room for skepticism, intracluster nebulosity and a 
short central cooling time are clearly connected.
All clusters known to have extended H$\alpha$ emission at 
their centers also have central cooling times shorter
than a Hubble time (Hu, Cowie, \& Wang 1985; Heckman \etal
1989; Baum 1992; Donahue 1997).  Moreover, the majority of 
clusters with short central cooling times contain such nebulae, and
their line luminosities correlate with $\dot{M}_X$.  

In hopes of nailing down this crucial piece of the cooling-flow
puzzle, astronomers have invented a variety of schemes to
link line emission to the cooling ICM.  The models proposed
have included repressurizing shocks (Cowie, Fabian, \& Nulsen 1980;
David, Bregman, \& Seab 1987), high-velocity shocks
(Binette, Dopita, \& Tuohy 1985), self-irradiated cooling 
condensations (Voit \& Donahue 1990; Donahue \& Voit 1991), and
turbulent mixing layers (Begelman \& Fabian 1990; Crawford \&
Fabian 1992).  None have been entirely successful.
Recent searches for [Fe~X] 6374~\AA~ emission from cooling-flow
clusters now seem to rule out any mechanism that relies on the
ionizing photons from cooling of hot gas (Donahue \& Stocke 1994;
Yan \& Cohen 1996). Photoionization by a central source also 
appears unlikely because the ionization level in 
cooling-flow nebulae remains constant while the pressure drops 
like $1/r$ (Johnstone \& Fabian 1988; Heckman \etal 1989).

Johnstone, Fabian, \& Nulsen (1987), motivated by the 
anticorrelation they found between the 
strength of the 4000~\AA~ break and the
H$\beta$ luminosities of cooling flows, proposed that hot stars
forming at the centers of cooling flows might photoionize
cooling flow nebulae.  Initially, this idea did not look
promising because the emission-line spectra of cooling-flow
nebulae look very different from H~II region spectra.
However, some recent work has strengthened the link between
emission lines and star formation in cooling-flow
clusters.  Unpolarized excess
blue continuum emission correlates with both the spatial
distributions aand luminosities of cooling-flow nebulae
(Allen 1995; Cardiel, Gorgas, \& Aragon-Salamanca 
1995, 1997; McNamara \etal 1996a,b).

In the meantime, we have obtained deep optical spectra of
cooling flow clusters that also implicate hot stars 
as the main stimuli of intracluster H$\alpha$.
Here we present our analysis of the emission lines
from Abell~2597.  Section~2 describes the observations,
and \S~3 discusses them.  Because we can detect a large
number of important optical lines, we can determine the
density, temperature, and metallicity at the center of the
emission-line
nebula without resorting to model-dependent assumptions.
Our analysis rules out both shocks and photoionization by a 
hard continuum, leaving hot stars as the most plausible
option.  However, hot stars have difficulty accounting
for the elevated temperatures we measure. Section~4
discusses why an additional source of heating might
be necessary and suggests that some sort of energy transfer
from the confining hot gas might well supply it.
Section~5 summarizes our results.

\section{Observations}

We observed Abell 2597 on August 12, 1993, in nearly photometric
conditions with the 5m Hale Telescope at the Palomar Observatory. 
Using the Double Spectrograph, we gathered both blue 
and red spectra simultaneously.  In these observations, a dichroic 
split the spectrum at approximately 5500 \AA, directing the blue light 
(3770-5500\AA) onto a grating with 300 lines/mm, a dispersion 
of 2.15 \AA/pixel, and an effective resolution of $\sim5$\AA, 
and the red light (5550\AA - 7980\AA) onto a grating with
316 lines/mm, a wavelength scale of 3.06 \AA/pixel, and an effective
resolution of $\sim 7-8$\AA.  Our three 3 exposures of Abell~2597
totalled 1.5 hours, and in each exposure a 2\arcsec~slit was placed 
on the central galaxy of Abell 2597 at a position angle of 30$^\circ$.  
The position angle remained within 30$^\circ$ 
of the parallactic angle throughout the observation. 

The data were processed in a standard way, using IRAF, in 
December 1993.  We removed pixel-to-pixel variations with
a normalized dome flat.  Our calibration of the wavelength
scale, using arc lamps of HC (blue) and NeAr (red) and some sky lines,
yielded wavelength solutions with 0.5\AA~RMS (red) and 1.2\AA~RMS (blue). 
Several exposures of a star at different positions along the slit 
were used to remove curvature along the dispersion axis. The IRAF task 
{\em fitcoords} in the {\em noao.twodspec.longslit} package 
was used to correct the data and the task {\em background} was 
used to select the source-free regions along the slit
and to subtract the sky contribution. The spectra were corrected for
atmospheric extinction and were flux corrected with standard 
star exposures of Feige 110 and BD33+2642, taken directly before 
the observations of Abell 2597.
Galactic extinction is predicted to be only 0.08 ($B$) magnitudes, 
so we did not deredden the spectrum. Any reddening we 
measure thus includes the effects of Galactic dust.

\begin{figure}
\plotone{\figdir 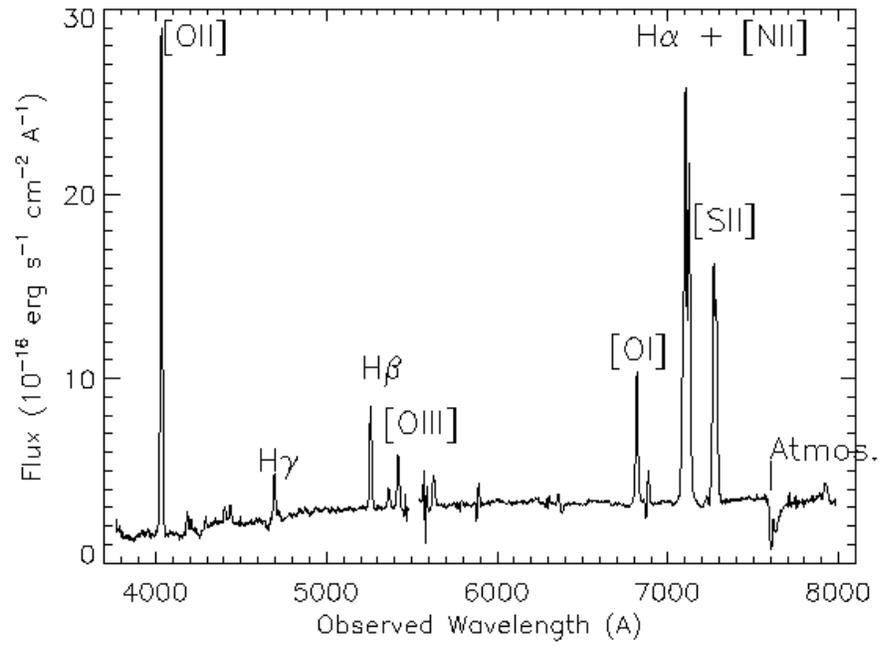}
\caption{Spectrum of the central galaxy of Abell~2597.
}
\label{all_a2597}
\end{figure}

\begin{figure}
\plotone{\figdir 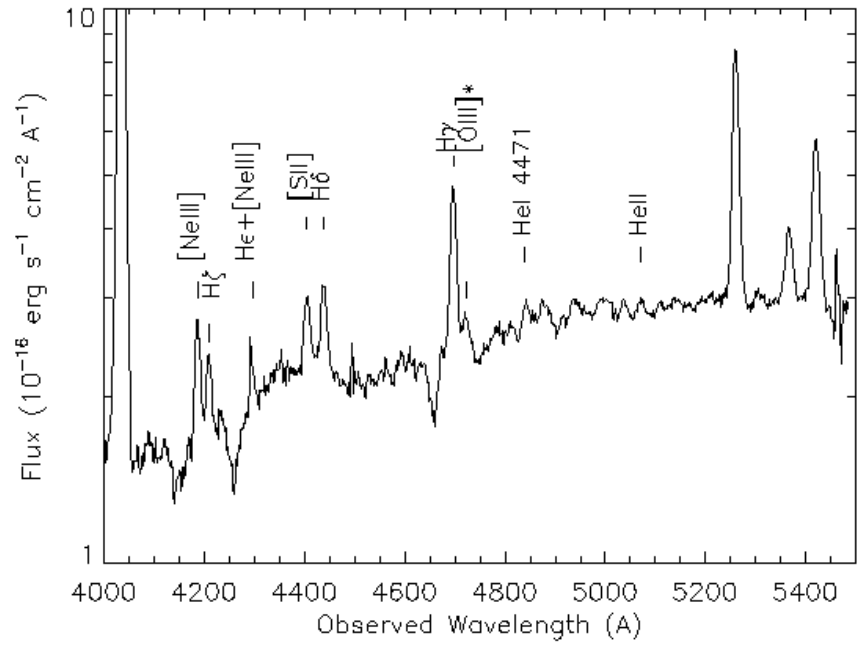}
\caption{Detailed view of blue spectrum of Abell~2597.
}
\label{b_a2597}
\end{figure}

\begin{figure}
\plotone{\figdir 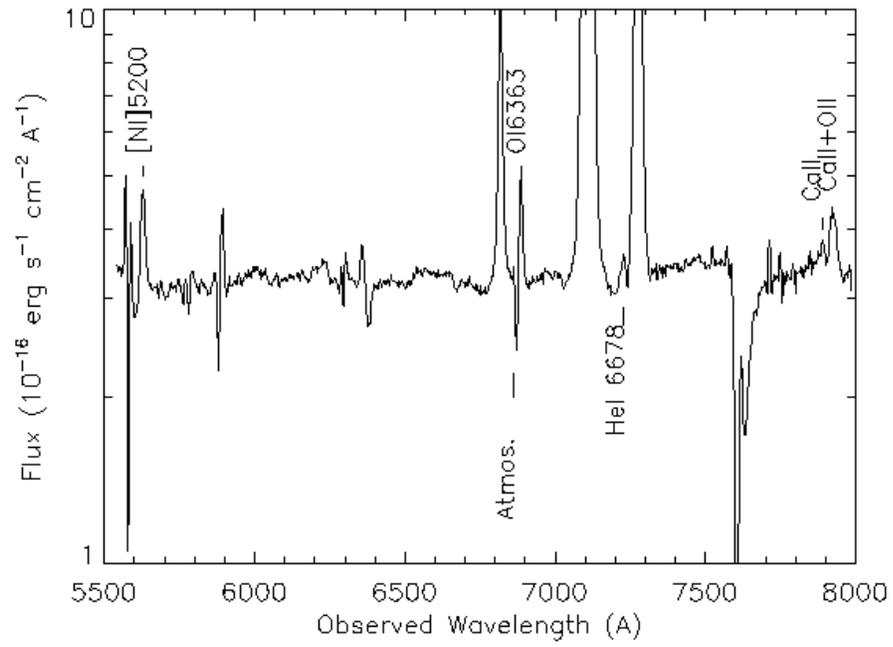}
\caption{Detailed view of the red spectrum of Abell~2597.
}
\label{r_a2597}
\end{figure}

We extracted a single spectrum, along 8 arcseconds 
of the slit centered on the nucleus, from each CCD 
(Figure~\ref{all_a2597}). The signal-to-noise ratio of this
spectrum exceeded 100 in the continuum and proved suitable
for studying the faint emission lines of interest
(Figures~\ref{b_a2597} and \ref{r_a2597}).
The emission line fluxes, measured using the {\em splot} task, 
are listed in Table~1. Errors in the fluxes of the red emission 
lines were assessed automatically by {\em splot}, which 
generates Monte Carlo simulations of the data and computes a 
1$\sigma$ (68.3\%) confidence range. The quoted errors in
the red lines thus include uncertainties in the background 
subtraction and Poissonian noise.

\begin{figure}
\plotone{\figdir 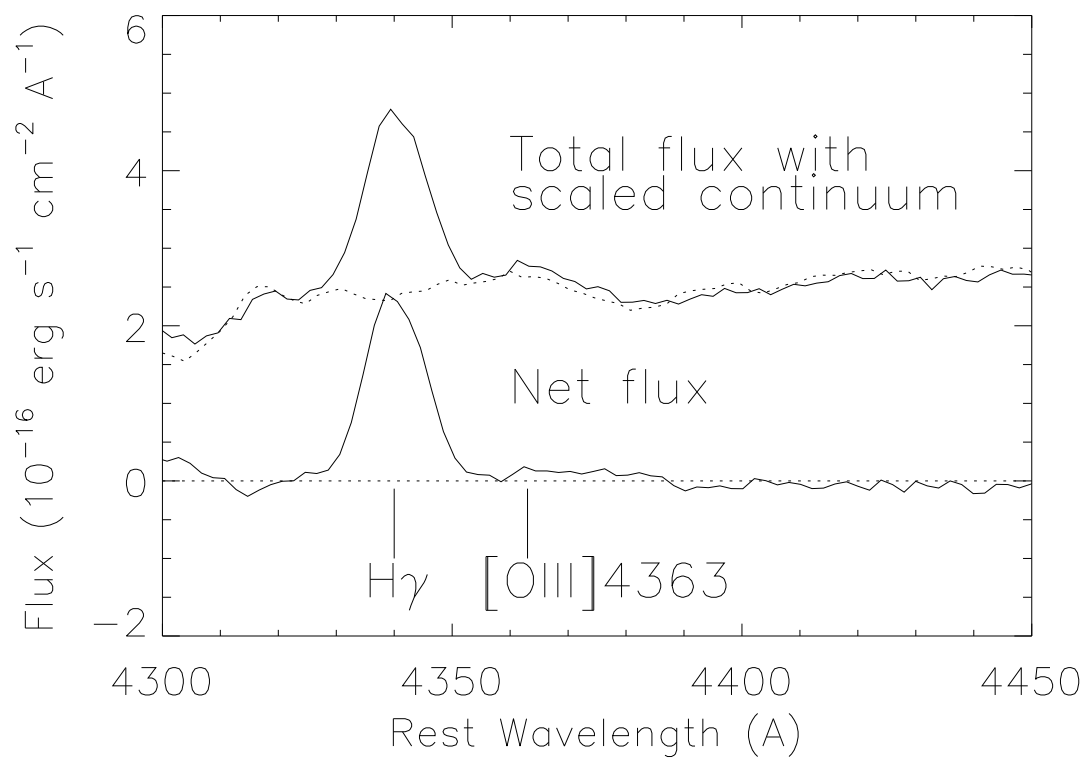}
\caption{Spectrum of Abell~2597 and template near [O~III] 4363.
Before template subtraction, a significant feature appears in
the total flux (upper solid line) near the expected position of 
[O~III] 4363.  After subtraction of the template (upper dotted line),
the feature disappears.  In the net flux (lower solid line), 
the 3$\sigma$ upper limit on the [O~III] 4363 line flux is
$1.3 \times 10^{-16} \, {\rm erg \, cm^{-3} \, s^{-1}}$.
}
\label{fig4363}
\end{figure}

\begin{figure}
\plotone{\figdir 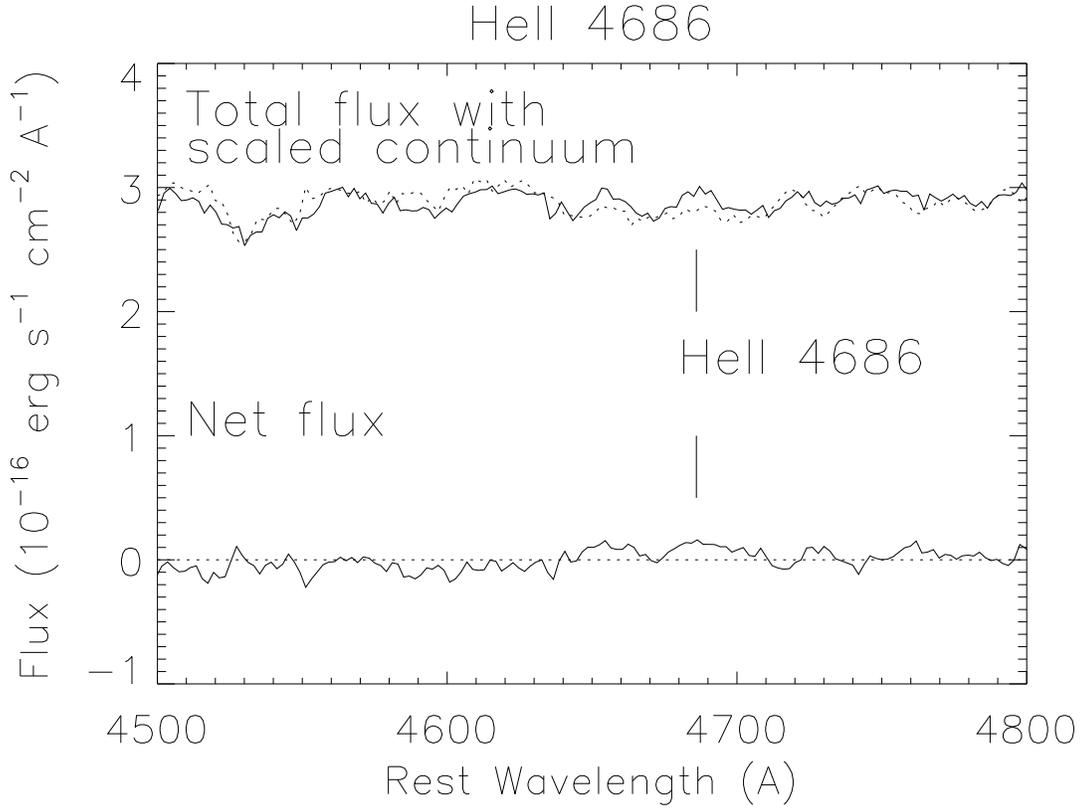}
\caption{Spectrum of Abell~2597 and template near He~II 4686.
The stellar continuum near the He~II 4686 line shows significant
structure.  After template subtraction, some net flux still
remains, producing a marginal feature whose flux, integrated
over the typical width of the other lines, has a formal significance 
of 6$\sigma$.  To the left of He~II, at a rest wavelength of 4655~\AA~
sits an unidentified feature whose integrated flux reaches 
a formal value of 4$\sigma$.
}
\label{fig4686}
\end{figure}

The complexity of the stellar continuum underlying the
blue lines in our spectrum necessitated a more
detailed line-measuring procedure.
To measure the blue lines, we constructed a template spectrum 
of nearby dwarf elliptical galaxies without emission lines. 
These spectra were acquired during the same night with 
the same instrumental setup. 
The absorption features in the template spectrum match
the absorption features in the A2597 spectrum quite well. 
We fitted the ratio of the A2597 spectrum to the template 
spectrum with a low-order polynomial, 
used this polynomial to scale the template spectrum, and
subtracted the scaled template from the A2597 spectrum.
Template subtraction proved to be very important. 
The A2597 emission line + stellar continuum spectrum showed 
a possible feature at the expected position of [O~III] $\lambda$4363, 
but this feature vanished upon subtraction of the stellar continuum 
because it was an artifact of the stellar absorption lines
(Figure~\ref{fig4363}).  Measuring the He~II 4686\AA~ recombination
line also depends critically on template subtraction 
(Figure~\ref{fig4686}).

In general, most of the error in the blue line measurements
stems from uncertainties in the subtraction of the background continuum. 
To gauge this uncertainty, we measured the lines 
manually many times, evaluating how different methods of background
removal changed the total line fluxes. Line flux estimates 
generally stayed within 5\% from measurement to measurement. 
For the weakest lines, this variation sometimes approached
10\%. Our line-flux uncertainties reflect this variation.  
Upper limits for the undetected lines were evaluated by measuring 
the residual RMS variation after subtraction of the stellar 
continuum template. For a FWHM=10\AA, the $3\sigma$ upper limit 
is $1.1 \times 10^{-16} \, {\rm erg \, cm^{-2} \, s^{-1}}$.
HeI 4471 was detected just barely above the 3$\sigma$ limit 
of $1.2-1.5 \times 10^{-16} \, {\rm erg \, cm^{-2} \, s^{-1}}$,
and we will treat this flux as an upper limit.

The lines all had a FWHM consistent with a velocity dispersion 
($\sigma$) of 270 km/sec (FWHM$=2.355\sigma$). The best-fit 
redshift is $0.0821 \pm 0.0002$.  No correction for the LSR 
was required because $V_{\rm LSR} = -13.5 \, {\rm km \, s^{-1}}$, 
smaller than our resolution.

\section{Emission-Line Analysis}

Most emission-line studies of cooling-flow nebulae have relied on accurate 
measurements of the strongest lines.  The relative fluxes of features like
H$\alpha$, H$\beta$, [N~II] $\lambda$6584, and [O~III] $\lambda$5007 
are generally handy for broad classification
of extragalactic nebulae, but by themselves they are not quite so 
useful for measuring physical quantities such as electron temperature 
and metallicity. With information on only a small set of lines,
we are usually left having to fit underconstrained photoionization 
models to the data.

This dataset on Abell~2597 enables us to do much more.  The extensive
set of line fluxes in Table~1 can be used to measure the reddening
of the emission-line spectrum, the density and temperature of the 
nebular gas, and the nebula's approximate metallicity, all without
recourse to underconstrained photoionization models.  We can also
put strong limits on the shock-excited [O~III] line at 4363~\AA~and
the He~II recombination line at 4686~\AA, measurements that rule out 
shocks as a major ionizing mechanism and point towards hot stars 
as the primary ionizing agent in the intracluster nebula.  

\subsection{Oxygen Lines \& Shocks}

One of our primary motivations in obtaining a deep blue 
spectrum of Abell 2597 was to measure the 4363~\AA~ emission 
line of [O~III].  Shocks radiate this line much more efficiently 
than photoionized gas, making it an effective diagnostic for 
distinguishing between these ionization processes.
Most shock models predict
\begin{equation}
    \ROIII  \equiv \frac {F_{4363}} {F_{4959} + F_{5007}}
                    \approx 0.05 - 0.07
\end{equation}
for the ratio between the [O~III] 4363~\AA~line and the sum of the 
4959~\AA~and 5007~\AA~lines.  
Photoionization models predict smaller $R_{\rm O \, III}$ values
because the [O~III] emitting gas is cooler.  At a temperature
of 10,000~K, $R_{\rm O \, III} \approx 0.005$ in low-density gas.
We find that at the center of Abell 2597, $R_{\rm O \, III} < 
0.02 \, (3\sigma)$, 
strongly indicating that shocks are not the dominant ionization
process there.

\begin{figure}
\plotone{\figdir 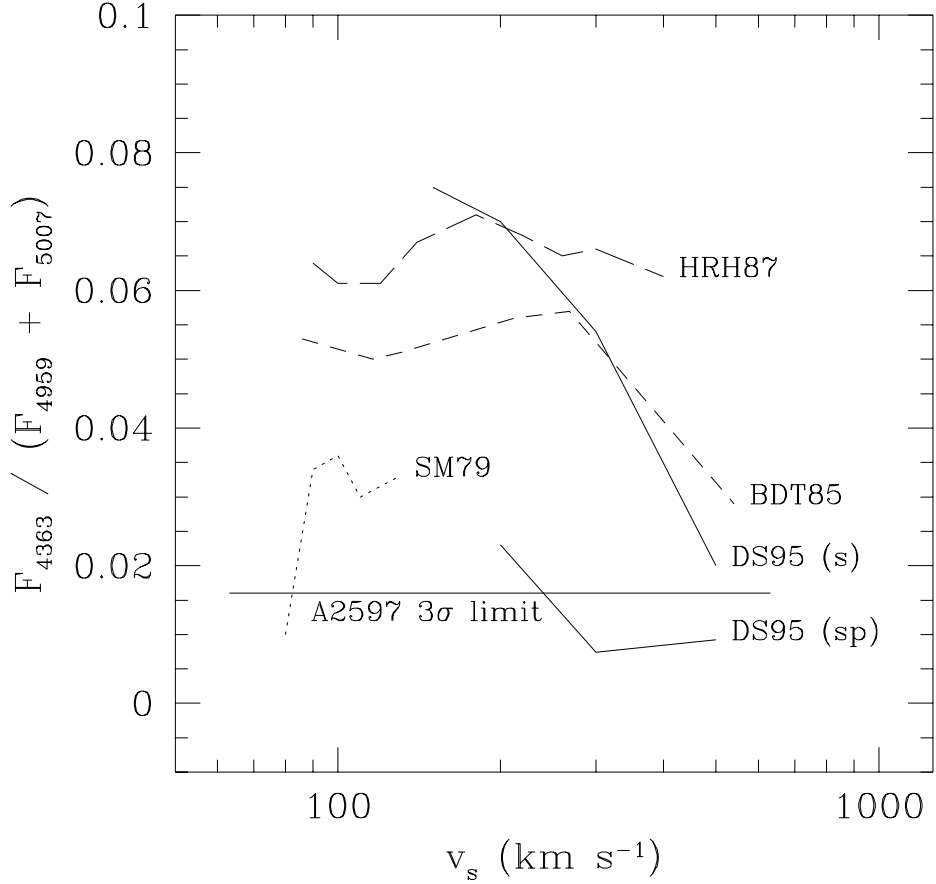}
\caption{[O~III] line ratios.  The temperature-sensitive [O~III]
line ratio $\ROIII \equiv F_{4363} / (F_{5007} + F_{4959})$ can 
distinguish between shocked gas and photoionized gas.  Gas behind 
shock fronts tends to be hotter than photoionized gas, leading to
larger $\ROIII$ values, if its density does not exceed $10^5 \, {\rm 
cm^{-3}}$.  This figure compares the $\ROIII$ predictions from several
shock models with our 3$\sigma$ upper limit.  The predicted ratios
fall below our upper limit only in shocks slower than 85~km~s$^{-1}$
and high-velocity shocks with photoionized precursors. 
(SM79 = Shull \& McKee 1979; 
BDT85 = Binette \etal 1985; 
HRH87 = Hartigan \etal 1987;
DS95 (s) = Dopita \& Sutherland 1995, shock only; 
DS95 (sp) = Dopita \& Sutherland 1995, shock+precursor).
}
\label{o3rat}
\end{figure}

\begin{figure}
\plotone{\figdir 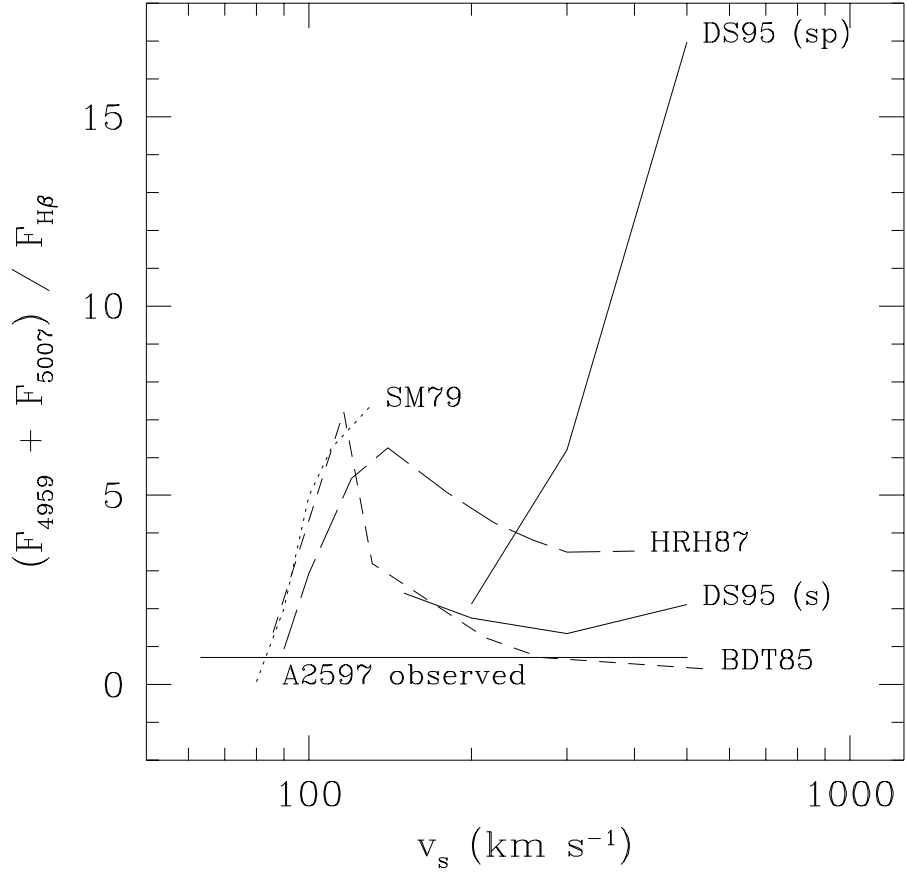}
\caption{[O~III] 4959 + 5007 / H$\beta$ line ratios.  This figure compares 
the observed [O~III] 4959 + 5007 / H$\beta$ line ratio with the ratios
predicted by shock models. Note that at high velocities in the DS95 (sp) 
model, where photoionization of the precursor reproduces $\ROIII$, the
expected [O~III] 4959 + 5007 / H$\beta$ ratios greatly exceed the
observed value.   Shock models reproduce both this line ratio and 
$\ROIII$ only in a very narrow range of shock velocities near 
80~km~s$^{-1}$.  However, shocks with these velocities fail to reproduce
the other forbidden-line ratios.
}
\label{o3_hb_rat}
\end{figure}

Figure~\ref{o3rat} shows $R_{\rm O \, III}$ predictions for shocks of
various velocities from the models of Shull \& McKee 
(1979; SM79), Binette \etal (1985; BDT85), Hartigan, Raymond, \& 
Hartmann (1987; HRH87), and Dopita \& Sutherland (DS95). 
Figure~\ref{o3_hb_rat} gives the corresponding ratios of [O~III] to H$\beta$.
As the shock velocity $v_s$ rises through $80 \, {\rm km \, s^{-1}}$,
the shock begins moving fast enough to collisionally ionize O$^+$ to
O$^{++}$, so the [O~III] fluxes rise rapidly.  In the cooling
postshock gas, the mean temperature of the [O~III] emitting gas 
does not vary much with shock velocity, and the $R_{\rm O \, III}$ 
ratio remains close to 0.05 up to shock velocities of
a couple hundred km~s$^{-1}$.

At higher shock velocities most of the [O~III] flux comes from
photoionized gas either upstream or downstream from the shock.
Ultraviolet radiation from the cooling gas behind the shock
propagates in both directions, ionizing whatever it encounters.
In the DS95 (s) and BDT95 models, which consider  
photoionization of only the downstream gas, $R_{\rm O \, III}$ 
almost drops below the observed limits, but the [Fe~X] limits 
rule out shocks this fast (Yan \& Cohen 1995).  The DS95 (sp) model 
also includes the photoionizing effects of postshock radiation 
on the lower-density upstream gas.  The photoionizing precursor 
creates a more highly ionized region that can radiate strongly 
in [O~III]. Such a shock satisfies the observed limits on 
$R_{\rm O \, III}$ but produces an [O~III]/H$\beta$ 
ratio far larger than observed.

In a very narrow range of shock velocities near 
80 km~s$^{-1}$ shock models come close to satisfying the
observed constraints on $R_{\rm O \, III}$ and [O~III]/H$\beta$
but fail for other reasons.
These shocks require extremely unlikely fine tuning and
do not reproduce the other forbidden-line ratios.
At 80 km~s$^{-1}$, the SM79 models predict [N~II]~6584/H$\beta
\approx 0.27$ and [S~II]~6717/H$\beta \approx 0.24$, whereas our
data show [N~II]~6584/H$\beta \approx 0.80$ and [S~II]~6717/H$\beta 
\approx 0.54$.  The disagreement of shock models with the
observed [O~III] lines leaves photoionization as the most
likely mechanism for generating the emission lines in Abell 2597.

\subsection{Balmer Lines \& Reddening}

Our spectrum of Abell 2597 contains hydrogen Balmer recombination
lines ranging from H$\alpha$ to H$\zeta$.  Assuming Case B 
recombination applies, we can use the Balmer series to measure 
the reddening of the visible spectrum.  The Balmer-line ratios 
decline systematically from red to blue, relative to Case B 
expectations at a density of $10^2 \, {\rm cm}^{-2}$ and a
temperature of 10,000~K, indicating that significant amounts of dust
obscure our view of the nebula.  The Galactic H~I column 
towards Abell~2597 is a modest $2.5 \times 10^{20} \, 
{\rm cm}^{-2}$, so the large majority of the 
obscuring dust must reside in the cluster.

\begin{figure}
\plotone{\figdir 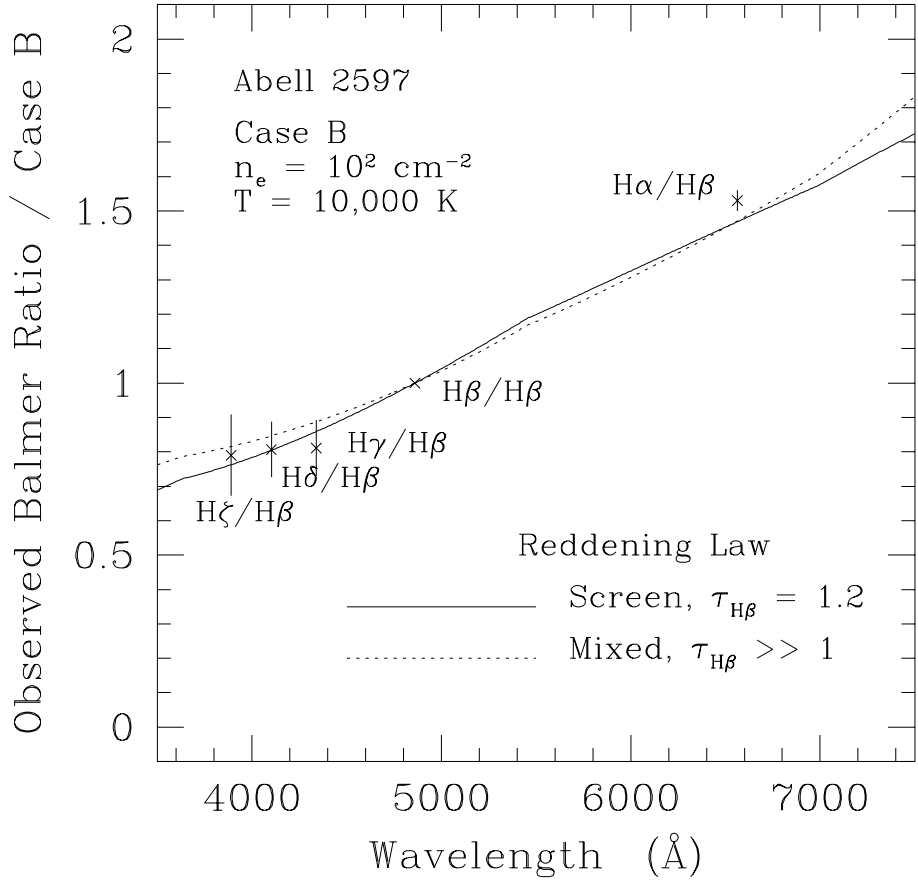}
\caption{Deviation of Balmer-line ratios from Case B. The Balmer-line
ratios in Abell~2597 are systematically redder than the Case~B predictions,
indicating significant reddening.  This figure shows how much these line
ratios deviate from their expected values in a plasma with $n_e = 10^2
\, {\rm cm^{-2}}$ and T = 10,000~K.  The solid line gives the
deviations produced by an obscuring screen of dust with an optical 
depth at H$\beta$ of 1.2, assuming a Galactic reddening law.  The 
dotted line shows the deviation expected if the line-emitting gas were
intermixed with dust having Galactic properties and a large overall
optical depth.  Our data cannot distinguish between these two dust 
distributions.
}
\label{redden_errs}
\end{figure}

Figure~\ref{redden_errs} illustrates how the inferred amount of dust depends
on the geometry of the dusty gas. To convert reddening to 
extinction, we assume a Galactic reddening law (Fitzpatrick 1986).
If the dust forms a screen interposed between us and the nebula,
the optical depth of the screen is $\approx 1.2$ at 
the wavelength of H$\beta$ ($A_V \sim 1$), corresponding 
to $N_{\rm H \, I} \approx 2 \times 10^{21} \, {\rm cm}^{-2}$
for a Galactic dust-to-gas ratio (Draine \& Lee 1984).  
The observed 
reddening of the Balmer lines turns out to be quite consistent 
with the properties of Galactic dust.  Alternatively, the grains 
could be intermixed with the line-emitting gas.  This scenario
requires a significantly larger dust column, one we cannot
limit from above.  The expected reddening in the large $N_{\rm
H \, I}$ limit is essentially indistinguishable from that in the 
screen model.

\subsection{Density \& Ionization Parameter}

\begin{figure}
\plotone{\figdir 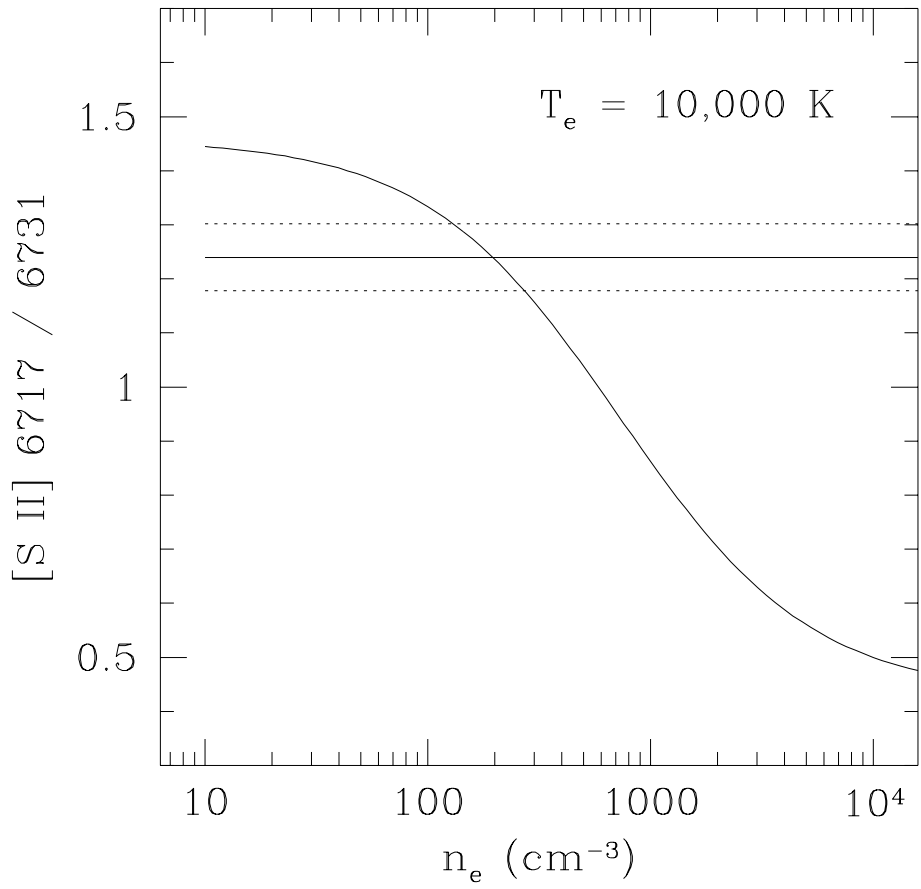}
\caption{Density of the cooling-flow nebula from the [S~II] ratio.
At the center of the nebula in Abell~2597, the [S~II] line ratio places 
the electron density at about $200 \, {\rm cm^{-3}}$.
}
\label{dens_new}
\end{figure}

The [S II] line ratio straightforwardly gives the electron density,
$n_e$, at the center of the nebula.  Figure~\ref{dens_new} shows that 
$n_e = 100 - 300 \, {\rm cm}^{-3}$, with a best value of
$200 \, {\rm cm}^{-3}$.  Such a density is typical of the inner 
regions of cooling-flow nebulae (e.g. Heckman \etal 1989).

We can combine this density measurement with the H$\alpha$ surface 
brightness to estimate the ionization level and column density
of the nebula.  The H$\alpha$ brightness at the center of the
nebula is $4.3 \times 10^{-15} \, {\rm erg \, cm^{-2} \, s^{-1}
\, arcsec^{-2}}$, implying an emission measure of $2100 \, {\rm
cm^{-6} \, pc}$.  At an electron density of $200 \, {\rm cm^{-3}}$, 
the inferred column density of ionized gas is then $N_{\rm H \, II} 
\approx 3 \times 10^{19} \, {\rm cm^{-2}}$.  Because the 21~cm 
absorption line in Abell~2597 indicates a much higher column density
of H~I, the ionized layers are likely to be thin skins on the surfaces
of thick neutral clouds (O'Dea, Gallimore, \& Baum 1994). 
The ionization parameter $U$ of a photoionized nebula is defined
to be the ionizing photon density divided by the number density of
hydrogen nuclei.  A thick photoionized slab of hydrogen gas illuminated
from one side has an ionized column $\sim (10^{23} \, {\rm cm}^{-2}) U$,
so the observed H~II column implies $U \sim 10^{-4}$, if the
nebulae are ionization-bounded and only a few distinct
photoionized surfaces lie along a given line of sight through the
nebula.  This value of $U$ is similar to that indicated by the
[O~III]/\hbeta ratio (Voit, Donahue, \& Slavin 1994).

\subsection{Temperature}

Our spectra contain temperature sensitive line sets from both
[O~II] (3726~\AA, 3729~\AA, 7320~\AA, 7330~\AA) and
[S~II] (4068~\AA, 4076~\AA, 6717~\AA, 6731~\AA). The two ratios
of interest are 
\begin{equation}
   \ROII \equiv \frac {F_{7320}+F_{7330}} {F_{3726}+F_{3729}}
\end{equation}
and
\begin{equation}
   \RSII \equiv \frac {F_{4068}+F_{4076}} {F_{6716}+F_{6731}} \; .
\end{equation}
To determine $\ROII$, we need to remove the contribution
of [Ca~II]~7324 to the red [O~II] line blend by subtracting
0.68 times the [Ca~II]~7291 flux.  Thus, $\ROII = 0.041 \pm 0.004$
and $\RSII = 0.042 \pm 0.004$.

\begin{figure}
\plotone{\figdir 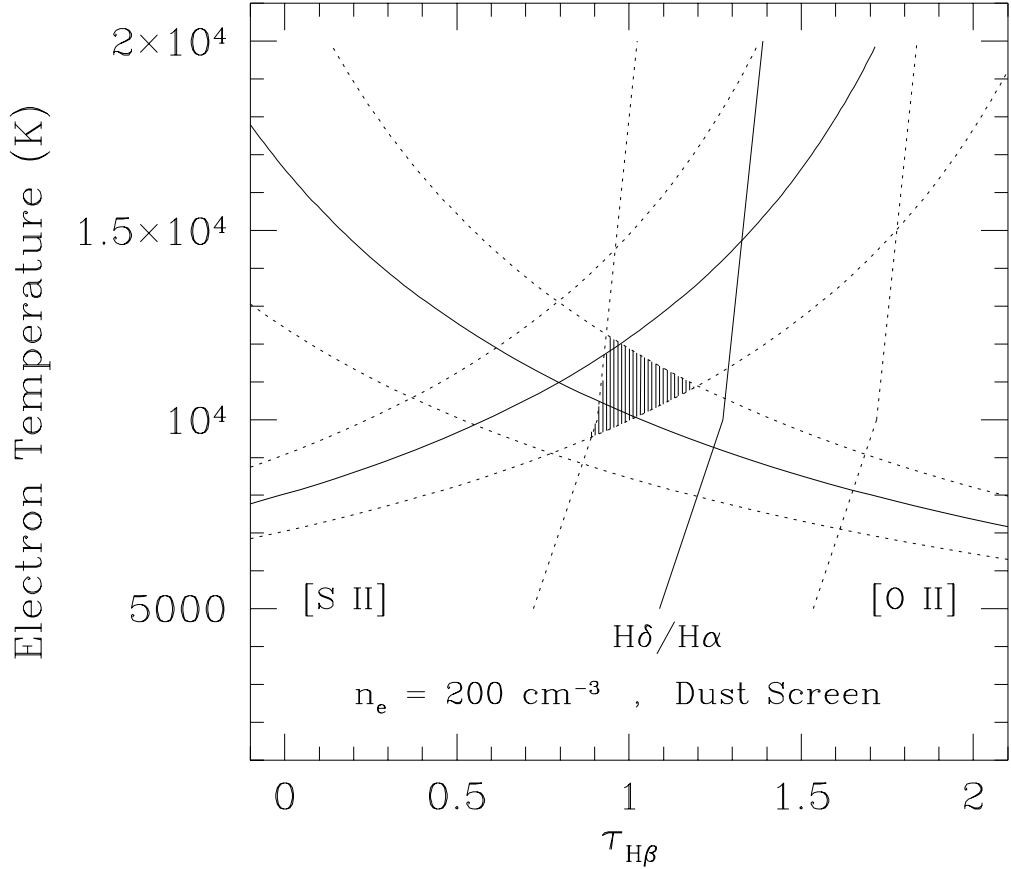}
\caption{Temperature constraints on the nebula in Abell~2597.
The temperature-sensitive [O~II] and [S~II] line ratios respond
to reddening in opposite ways, as shown by the solid lines. 
Jointly these line ratios constrain both the reddening and 
the temperature of the nebula.  The H$\delta$/H$\alpha$
ratio also constrains reddening and limits the permitted region
in the temperature-reddening plane even further.  Dotted lines
trace the 2$\sigma$ uncertainty ranges associated with each line
ratio.  Together, these three line ratios limit the electron
temperature to 9,500~K $<$ T $<$ 12,000~K.
}
\label{temps_shade}
\end{figure}

To measure accurate temperatures, we need to correct these 
line ratios for reddening.  Luckily, reddening affects the temperatures 
derived from these line ratios in opposite ways.  In the 
temperature-reddening plane, the true temperature should 
lie at the intersection of the loci defined by $\ROII$ and $\RSII$.
Figure~\ref{temps_shade} illustrates how the temperatures inferred from 
these ratios in Abell 2597 change as the dust optical depth
at H$\beta$ ($\tau_{{\rm H}\beta}$) increases, given an electron density
of 200~cm$^{-3}$ and a screen model for the obscuring dust.  
We also include the locus defined in this same plane by the 
H$\delta$/H$\alpha$ ratio.  Within the observational errors, 
these three loci all intersect in a region where
$9500 \, {\rm K} < T_e < 12,000 {\rm K}$. 

Note that these temperature limits do not depend significantly 
on the reddening model.  The red [S~II] lines are near H$\alpha$
and the blue [S~II] lines are near H$\delta$, so the necessary
Balmer-line correction gives the [S~II] reddening correction 
directly.  The exact correction for the [O~II] lines should differ
only slightly from the [S~II] correction.  Note also that 
the 9,500~K lower limit on the electron temperature does 
not depend on $\ROII$.

\subsection{Metallicity}

Because we know both the temperature and density of the
ionized gas, the ratio of each forbidden-line flux 
to the hydrogen Balmer
lines tells us the abundance of each line-emitting species. 
To derive elemental abundances from these lines, we would 
also need to know the ionization level in the plasma.
Given the small ionization parameter (\S~3.2), the ionization
levels are likely to be low, with most species predominantly
singly ionized.

\begin{figure}
\plotone{\figdir 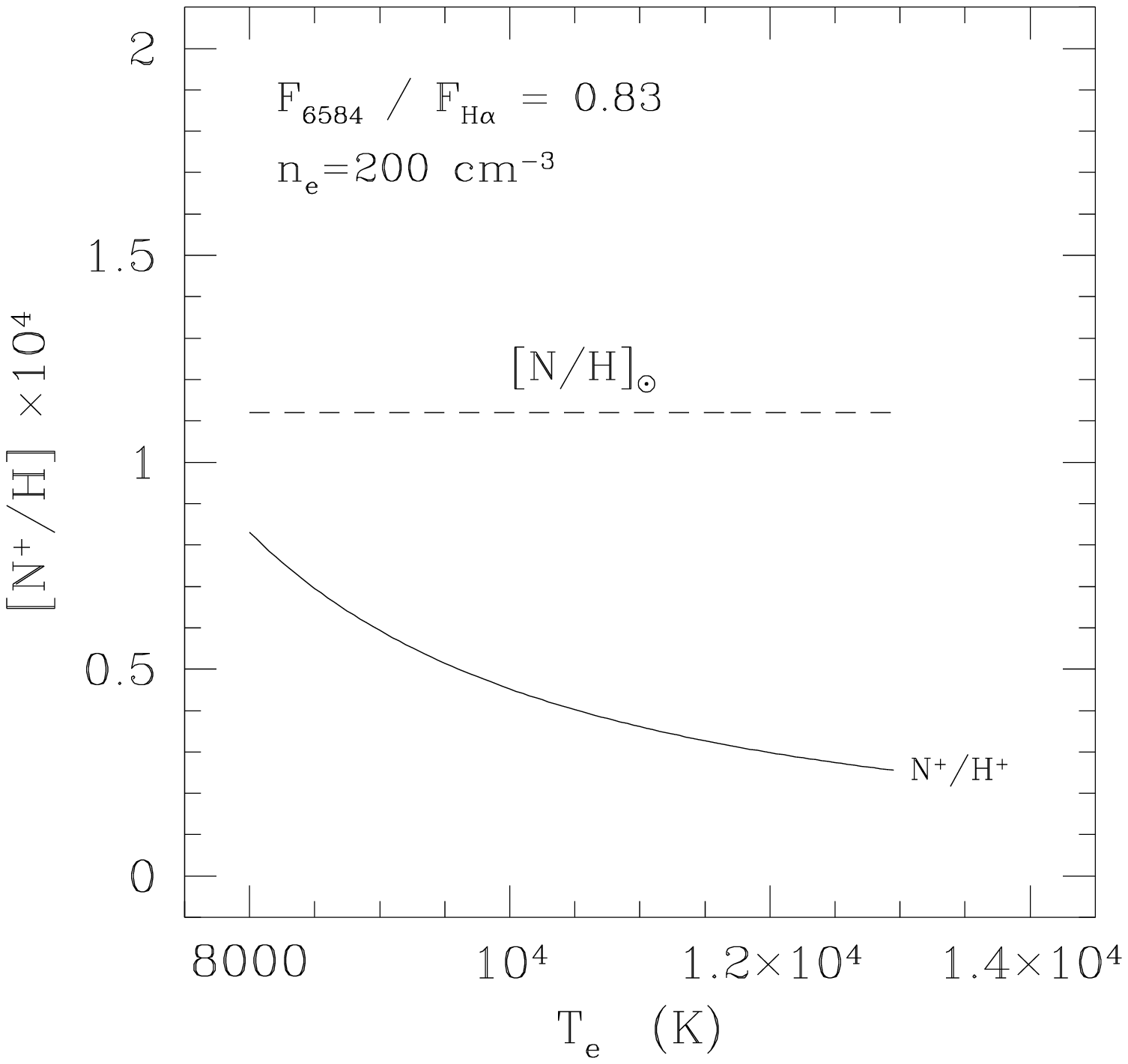}
\caption{N$^+$ abundance/temperature relation for Abell~2597.
The solid line shows the N$^+$/H$^+$ ratio derived from the
observed $F_{6584}/F_{{\rm H}\alpha}$ ratio for various assumed
electron temperatures ($T_e$) and an electron density of 
200~cm$^{-3}$.  The dashed line gives the solar N/H
abundance.
}
\label{nii_rat}
\end{figure}

Figure~\ref{nii_rat} shows how the derived abundance of N$^+$ varies
with electron temperature for the [N~II]~6584/H$\alpha$
line ratio of 0.83 observed in Abell 2597.  If the electron 
temperature is $\sim 9,500$~K, the N$^+$/H$^+$ ratio is about
half of the solar N/H ratio.  At the upper end of the allowed
temperature range, near 12,000~K, the derived N$^+$/H$^+$
ratio drops to about 1/4 of the solar N/H.

\begin{figure}
\plotone{\figdir 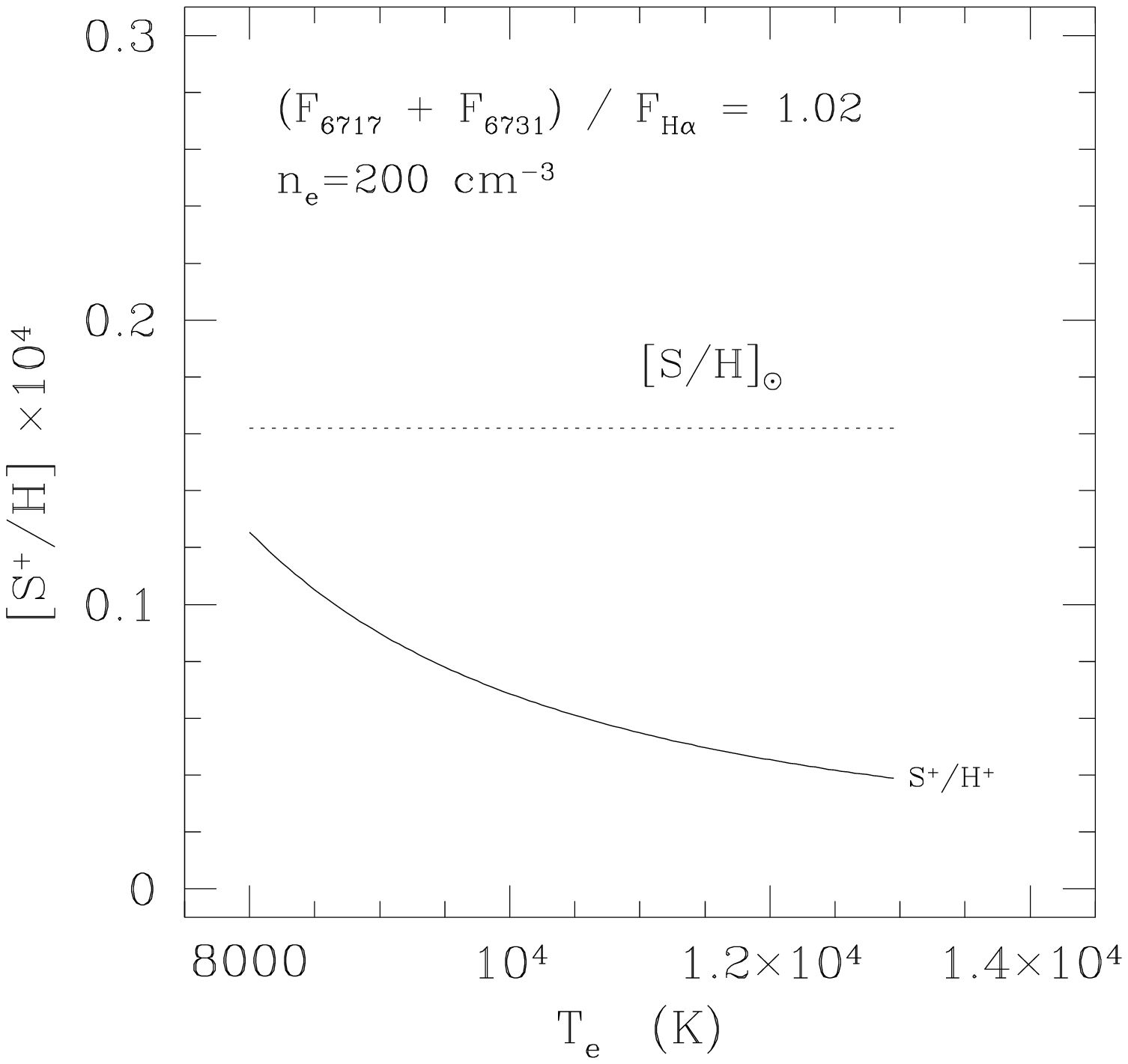}
\caption{S$^+$ abundance/temperature relation for Abell~2597.
The solid line shows the S$^+$/H$^+$ ratio derived from the
observed $(F_{6717}+F_{6731})/F_{{\rm H}\alpha}$ ratio for various assumed
electron temperatures ($T_e$) and an electron density of 
200~cm$^{-3}$.  The dashed line gives the solar S/H
abundance.
}
\label{sii_rat}
\end{figure}

The story for S$^+$ is similar to that for N$^+$.
Figure~\ref{sii_rat} shows the derived range of S$^+$/H$^+$, which
runs from half-solar at 9,500~K to 1/4 solar at 12,000~K.
Sulphur is somewhat more likely to be doubly ionized than
nitrogen, as the ionization potential for S$^+$ is 23.3~eV,
compared to 29.6~eV for N$^+$. Nevertheless, the close 
correspondence between the relative abundances of N$^+$ 
and S$^+$ suggests that sulphur is mostly singly-ionized.

\begin{figure}
\plotone{\figdir 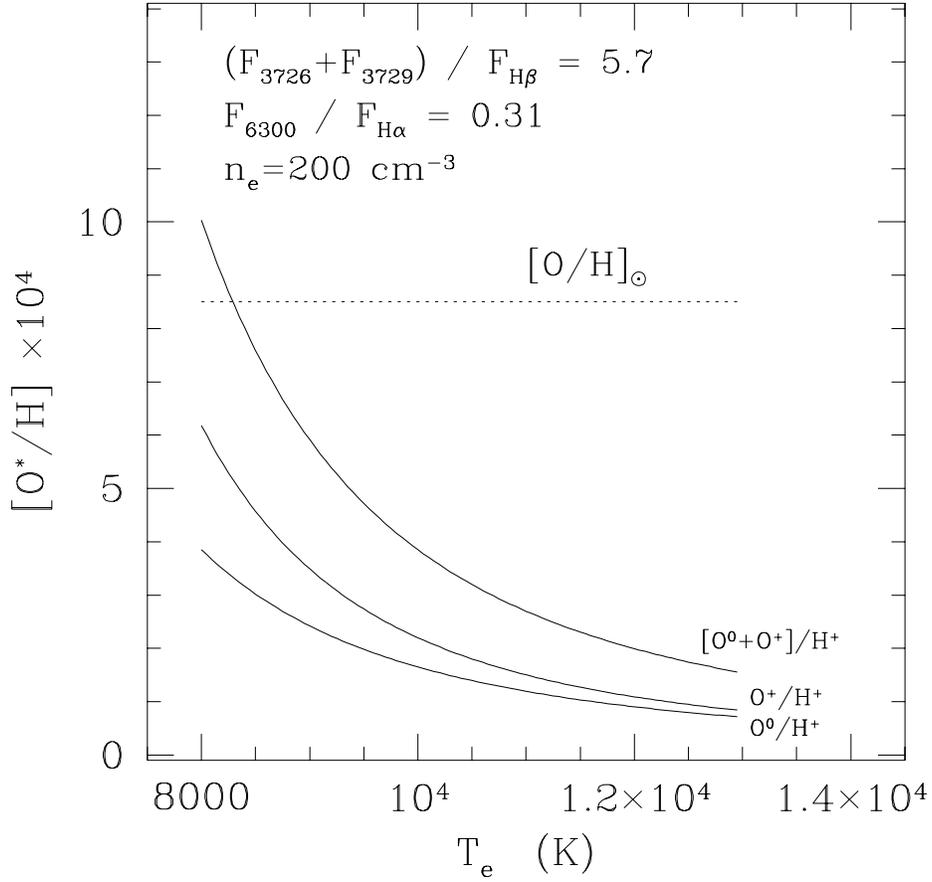}
\caption{O$^+$+O$^\circ$ abundance/temperature relation for Abell~2597.
The solid lines show the O$^+$/H$^+$, O$^\circ$/H$^+$, and 
[O$^\circ$+O$^+$]/H$^+$ ratios derived from the observed 
$(F_{3726}+F_{3729})/F_{{\rm H}\alpha}$ and 
$F_{6300}/F_{{\rm H}\alpha}$ ratios 
for various assumed electron temperatures ($T_e$) and an 
electron density of 200~cm$^{-3}$.  The dashed line gives the 
solar O/H abundance.  Because some of the [O~I] line flux
could be coming from neutral gas, the actual O$^\circ$/H
ratio in the ionized nebula might be lower.
}
\label{o_rat}
\end{figure}

Our determination of the oxygen abundance benefits from
lines radiated by three different ionization states.  
Figure~\ref{o_rat} shows the relative abundances of O$^+$ 
and O$^\circ$, with respect to H$^+$.  As with N and S,
their sum corresponds to half of the solar O/H ratio
at 9,500~K and to 1/4 the solar value at 12,000~K.
The relatively modest [O~III] lines indicate that O$^{++}$ 
is a minority species; both O$^+$ and O$^\circ$ are more common.
The consistency of these abundances with those for nitrogen
and sulphur is comforting; however, some of the [O~I] flux 
could be coming from X-ray heated neutral gas behind the 
ionized region, so the derived O$^\circ$/H$^+$ is really
an upper limit for the ionized gas.


\subsection{Helium Recombination Lines}

The forbidden lines in cooling-flow nebulae are generally
quite strong, relative to the Balmer recombination lines, 
indicating a large amount of heat input per photoionization.
One possible way to supply the requisite heating is with a hard
photoionizing continuum that remains strong into the soft
X-ray band.  Models that have used such an ionizing spectrum 
to reproduce some of the major line ratios include fast-shock 
models (Binette et al. 1985), self-irradiation models 
(Voit \& Donahue 1990, Donahue \& Voit 1991), and 
mixing-layer models (Crawford \& Fabian 1992).

The helium recombination lines provide a model-independent
way to check the hardness of the incident continuum.  Our
spectrum of Abell~2597 includes the He~II recombination
line at 4686~\AA~and the He~I recombination lines at
6678~\AA, 5876~\AA, and 4471~\AA.  Broad Na~I absorption,
possibly associated with the dusty H~I gas, contaminates 
the 5876~\AA~line, reducing its usefulness. The story told 
by the 4471~\AA~ line is also unclear because the underlying 
continuum is so noisy.  We barely detect this line at a level 
of $F_{4471}/\Hb \approx 0.02$, whereas we would expect 
$F_{4471}/\Hb \approx 0.05$ if all the helium were singly ionized.
The 6678~\AA~ line, located in a cleaner part of 
the spectrum, is probably the most reliable.
At the center of the cluster, $F_{6678}/F_{\Ha} = 0.01$, 
as expected in a nebula where singly-ionized helium 
predominates.  In contrast, $F_{4686}/F_{\Hb} \approx 0.02$,
implying very little He$^{++}$.

The lack of He~II recombination-line flux from Abell~2597
indicates that the photoionizing continuum cannot be 
especially hard.  If the incident spectrum follows a $F_\nu \propto
\nu^\alpha$ power law from 13.6~eV into the soft X-ray band,
then $\alpha \approx -2.3$ for $U \sim 10^{-4}$.  This kind 
of spectrum is too soft to supply the necessary heating per 
ionization.  The He~II/H~I line ratio argues against AGN 
irradiation, fast shocks, and cooling gas as sources of the 
photoionizing continuum, leaving hot stars as the only plausible 
possibility.\footnote{We note, however, that a significant minority 
of quasars do have He~II 4686/\hbeta ratios this small (e.g. Boroson \&
Green 1992).  The ionizing continuua of these objects must be far
weaker than the standard AGN continuum at the He~II edge and cannot 
sustain the high nebular temperatures we observe.}

\section{Energetics}

The line-ratio analysis of the previous section eliminated
many of the most frequently mentioned energy sources for 
cooling-flow nebulae.  Hot stars were not ruled out by
these model-independent diagnostics, but can we construct
a hot-star photoionization model that actually works?
In this section, we attempt to do so, but we find that
standard hot-star photoionization fails to 
provide the requisite heating by a factor of about two.
We speculate that some kind of energy transfer from the
hot surrounding medium might supply the other half of
the heating.

\subsection{Stellar Photoionization: Insufficient}

\begin{figure}
\plotone{\figdir 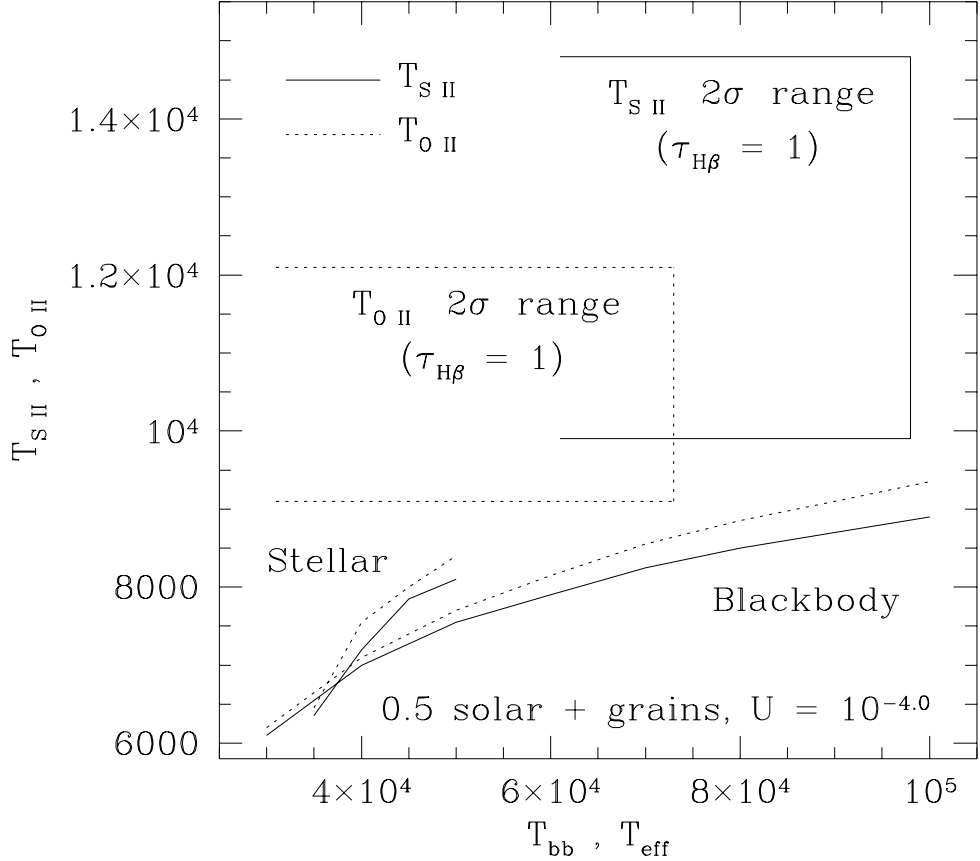}
\caption{Comparison between model H~II regions and temperatures 
measured from [S~II] and [O~II] line ratios.  The solid (S~II) and
dotted lines (O~II) give the temperatures inferred from the S~II
and O~II line ratios given by CLOUDY (Ferland 1993) photoionization 
models of H~II regions.  The input spectra for the `Stellar' models 
are Kurucz model atmospheres with effective temperatures ($T_{\rm eff}$)
ranging from 35,000~K to 50,000~K.  The lines labelled `Blackbody'
give the results for blackbody input spectra with temperatures
from 30,000~K to 100,000~K. In the photoionized gas, the abundances
are 0.5 solar with Galactic depletions, and the ionization parameter
is $\log U = -4.0$.  All the models fail to reproduce the observed 
temperatures.
}
\label{tmodels}
\end{figure}

To investigate whether hot stars alone can produce the
emission lines observed in Abell~2597, we constructed 
numerous models with the photoionization code CLOUDY, 
version 84 (Ferland 1993).  The code includes a grid of
Kurucz model atmospheres up to effective temperatures of
50,000~K, as well as blackbody spectra at any temperature. 
We tested stellar models with $T_{\rm eff} = 35,000 - 50,000$~K
and blackbody models with $T_{\rm bb} = 30,000 - 100,000$~K
incident upon a gas with 0.5 solar metallicity, Galactic
depletions, and $n_e = 200 \, {\rm cm^{-3}}$.
Figure~\ref{tmodels} compares the temperatures derived from
the [S~II] and [O~II] lines in these models with the temperature
ranges derived from the observed lines, 
assuming $\tau_{{\rm H}\beta} = 1$.  
Only in the very hottest blackbody models
do the electron temperatures approach the observed values.
Thus, the continuua of hot stars are not hard enough to produce
temperatures exceeding 9,500~K in gas with half-solar metallicity.
If stars are responsible for photoionizing the nebula in Abell~2597,
some other energy source must be providing additional heating
comparable in magnitude to the photoelectric heating.

\begin{figure}
\plotone{\figdir 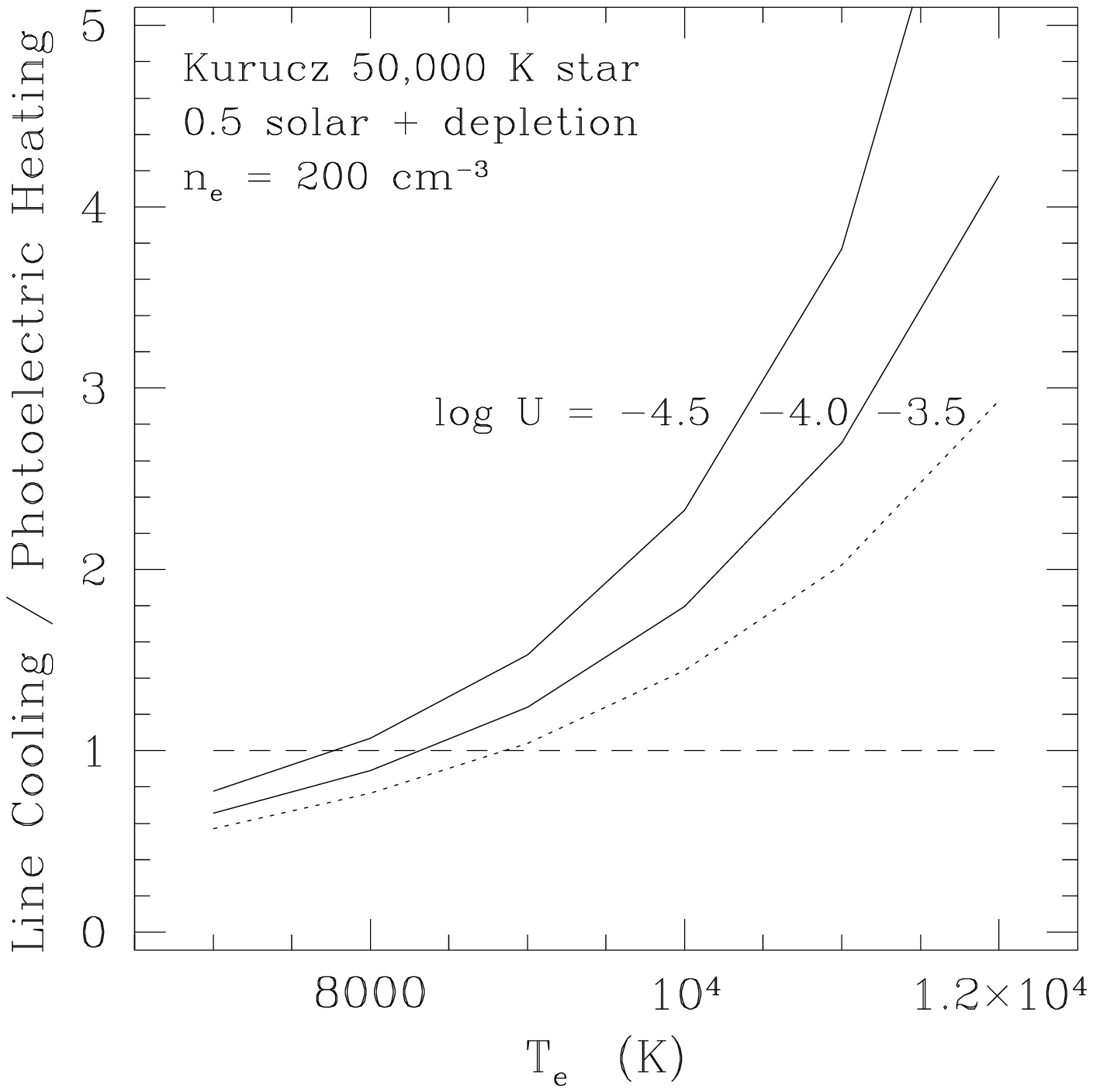}
\caption{Heating/Cooling budget in stellar H~II regions of 0.5 solar
metallicity.  The lines show the ratio between photoelectric heating
and line cooling in photoionized gas over a range of electron
temperatures ($T_e$).  A Kurucz model with $T_{\rm eff} = 50,000$~K
provided the ionizing spectrum, irradiating gas with 0.5 solar
abundance, Galactic interstellar depletions, and an electron
density of 200~cm$^{-3}$.  Two solid lines trace the ratio of line cooling
to photoelectric heating for $\log U = -4.5$ and 4.0.  The dotted
line traces the same ratio for $\log U = -3.5$, a value of $U$
that overproduces the [O~III] lines.  The intersection of the
dashed line with the others gives the equilibrium temperatures
of the models.
}
\label{heat50_50}
\end{figure}

Figure~\ref{heat50_50} shows a specific example of stellar
insufficiency.  In these models, a hot stellar continuum with
$T_{\rm eff} = 50,000$~K irradiates a gas with half-solar 
metallicity.  When the illuminating star is this hot, 
the O~III/\hbeta ratio constrains the ionization parameter
to be $U \approx 10^{-4.0} - 10^{-4.5}$, in agreement with our
estimate from the H$\alpha$ surface brightness (\S~3.3).  
Refractory elements are depleted into dust grains, 
as in our own interstellar medium, and photoelectric 
heating from dust contributes energy to the nebula.  
Both depletion of coolants and dust heating raise 
the expected equilibrium temperature.  
Even so, the equilibrium temperatures set by photoelectric
heating barely rise above 8000~K.  

The extra heating required to boost $T_e$ above 9,500~K is
comparable to the photoelectric heating itself.  
In equilibrium at these higher temperatures
the sum of photoelectric heating and some other supplementary 
form of heating must balance line cooling. 
To achieve temperatures in the observed range, the total 
heating must be roughly twice the photoelectric
heating.  Even when the ``extra heating'' is several times the 
photoelectric input, Ly$\alpha$ cooling keeps the equilibrium
temperatures below 12,000~K.  

\begin{figure}
\plotone{\figdir 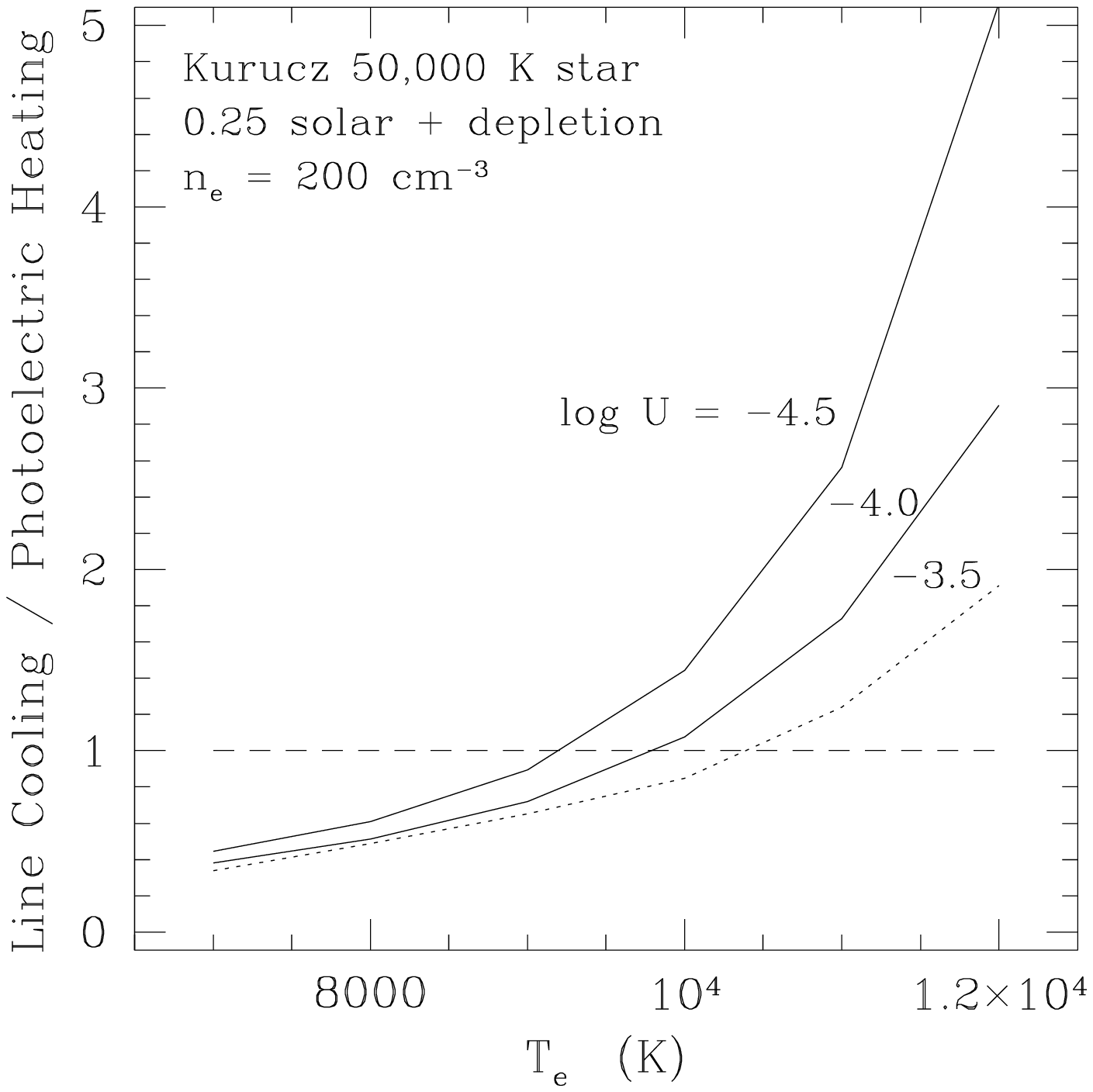}
\caption{Heating/Cooling budget in stellar H~II regions of 0.25 solar
metallicity.  The lines show the ratio between photoelectric heating
and line cooling in photoionized gas over a range of electron
temperatures ($T_e$).  A Kurucz model with $T_{\rm eff} = 50,000$~K
provided the ionizing spectrum, irradiating gas with 0.25 solar
abundance, Galactic interstellar depletions, and an electron
density of 200~cm$^{-3}$.  Two solid lines trace the ratio of line cooling
to photoelectric heating for $\log U = -4.5$ and 4.0.  The dotted
line traces the same ratio for $\log U = -3.5$, a value of $U$
that overproduces the [O~III] lines.  The intersection of the
dashed line with the others gives the equilibrium temperatures
of the models.
}
\label{heat50_25}
\end{figure}

Lowering the metallicity does not fix the mismatch between 
photoelectric heating and line cooling.  In \S~3.5 we showed that 
metallicities around 1/4 solar are possible if $T_e \approx 
12,000$~K.  Figure~\ref{heat50_25} illustrates that stellar
photoionization indeed generates higher equilibrium temperatures
in lower metallicity gas, but the extra heating needed 
to boost $T_e$ all the way up to 12,000~K is still $2-4$ times the
photoelectric heating.  Models with half-solar metallicities
actually require smaller proportions of additional heating.

\subsection{Extra Heating}

The failure of stellar photoionization to account fully for
the heating of the nebula might be surprising, given the
convincing evidence in favor of stellar photoionization as 
the primary ionizing mechanism.
On the other hand, another ample source of heat energy, namely
the pervasive hot intracluster gas, lies close at hand and presumably
confines the ionized surfaces of the nebula.  Several authors have 
pointed out that the energy fluxes necessary to power the most luminous 
cooling-flow nebulae are similar to the product of the pressure 
$P$ and the sound speed $v_{\rm th}$ in the surrounding medium 
(Heckman \etal 1989; Donahue \& Voit 1991; Crawford \& Fabian 1992).  
Here we argue that this property of cooling-flow nebulae 
may not be entirely coincidental.

Energy can conceivably flow from the hot ICM into the cooler
nebula in numerous ways.  A few specific examples are electron 
thermal conduction (e.g. Sparks 1992), acoustic wave heating
(e.g. Pringle 1989), and MHD wave heating (e.g. Friaca \etal 1997).
All of these heat fluxes saturate at a rate $\sim Pv_{\rm th}$.

Assume for the moment that thermal energy passes from the hot medium
into the photoionized surfaces of the nebula at the saturated flux
\begin{equation} 
  F_{\rm sat} = f_{\rm sat} P v_{\rm th} \; \; ,
\end{equation} 
where $f_{\rm sat}$ is a parameter of order unity.  Meanwhile, the 
flux of photoelectric heat energy into the nebula is 
\begin{equation}
  F_{\rm ph} \approx \frac {PcU(\bar{E}_{\rm ph} - 13.6 \, {\rm eV})} 
                           {2.3 k(10,000 \, {\rm K})}
\end{equation}
where $\bar{E}_{\rm ph}$ is the mean energy per ionizing photon.
The ratio of these energy fluxes is
\begin{equation}
  \frac {F_{\rm sat}} 
        {F_{\rm ph}} = 1.5 \, f_{\rm sat} 
                       \left( \frac {U} {10^{-4}} \right)^{-1}
                       \left( \frac {v_{\rm th}} {300 \, {\rm km s^{-1}}} \right)
                       \left( \frac {\bar{E}_{\rm ph}} {13.6 \, {\rm eV}}
                               - 1 \right)^{-1} \; \; .
\end{equation}
Note that in normal H~II regions, with $U \sim 10^{-2} - 10^{-3}$, 
the photoelectric heat flux overwhelms any mechanism operating 
at the rate $F_{\rm sat}$.  Such processes become energetically
significant only where $U$ is low, in nebulae with dilute radiation
fields or high pressures.  At the low ionization parameters 
observed in Abell~2597 and other cooling-flow nebulae, a saturated 
mechanical energy flux could easily be comparable to the photon 
energy flux. 

While these alternative mechanisms for supplying ``extra heating'' 
appear promising, and maybe even inevitable, on the broad level 
of the energy budget, further work is required to see if they will
dissipate and distribute their energy throughout the nebula.  
For instance, saturated electron thermal conduction can transfer 
energy at a maximum flux of
\begin{equation}
  F_{e^-} \approx 5 P v_{\rm th}
\end{equation}
(Cowie \& McKee 1977), but
the penetration depths of hot electrons into the nebula
depend strongly on their velocities.  An electron of
energy $T_{\rm keV} (1 \, {\rm keV})$ passing into a fully ionized
$10^4$~K plasma of density $n_e = 200 \, {\rm cm^{-3}}$ will stop after 
traversing a column density
\begin{equation}
  N_{\rm stop} \approx (2 \times 10^{17} \, {\rm cm^{-2}}) T_{\rm keV}^2
\end{equation}
(Voit 1991).  Comparing $N_{\rm stop}$ to the typical thickness of 
a low-ionization H~II layer, $\sim (10^{19} \, {\rm cm^{-2}})(U/10^{-4})$,
shows that the distribution of conductive heating in
the nebula will depend sensitively on the exterior temperature.
However, electrons at the expected energies of several keV are 
clearly energetic enough to penetrate most of the ionized layer, 
as long as the magnetic field geometry is favorable.  
A coupled photoionization/conduction code will be needed to solve 
for the actual temperature structure and line emission from such 
a cloud.

\subsection{LINERs in General}

Classifiers of extragalactic nebulae place cooling-flow nebulae
among the LINERs, a heterogeneous class of objects whose [O~II]
$\lambda$3727 lines are stronger than their [O~III] $\lambda$5007
lines and whose [O~I] $\lambda$6300 lines are greater than 1/3
of [O~III] $\lambda$5007 (see Filippenko 1996 for a recent review).
Generally, LINERs also have [N~II] $\lambda$6584 / \halpha $>$ 0.6, 
unusually high for extragalactic H~II regions (e.g. Ho 1996).
These large forbidden-line fluxes, relative to \halpha and \hbeta,
imply high electron temperatures difficult to attain
with normal O-star photoionization.  While photoionization 
by an AGN-like nuclear X-ray source might explain the line ratios
in compact LINERs, the line emission in a significant fraction 
of LINERs extends over a few kiloparsecs and appears to be powered
by local processes (Filippenko 1996).  Furthermore, the He~II 
$\lambda$4686/\hbeta ratios in LINERs are frequently lower than
AGN-like models would predict (Netzer 1990).

Quite possibly, supplementary heating of the kind proposed here 
for cooling-flow nebulae could account for the strong forbidden lines 
of LINERs in general, regardless of the ionizing source.
Mechanical and conductive forms of heating probably operate 
constantly at some level in all kinds of photoionized nebulae.
Usually, the photon fluxes incident on these nebulae overwhelm
any other energy source.  However, when the ambient pressure rises
or the photon flux decreases, the ionization parameter drops, 
reducing the dominance of photoelectric heating.  If mechanical 
or conductive processes transfer heat from a hot confining medium
to photoionized clouds at the saturation rate ($\sim P v_{\rm th}$), 
then we would expect their contributions to become significant 
when $U \lesssim 10^{-3.5}$, the ionization range characteristic 
of LINERs. 
A variety of physical processes, differing from one astrophysical
environment to another, could conspire to produce this characteristic
combination of low ionization and supplementary heating 
in many different situations.  The
spectral signatures of these objects would all look LINER-like,
even though the underlying phenomena differ.  Their only shared
property would be a value of $U$ low enough for other forms of 
heating to augment photoelectric heating.

\section{Summary}

Our deep spectra of the cooling-flow nebula in Abell~2597 
strongly constrain the possible ionizing mechanisms.  The
lack of [O~III] $\lambda$4363 emission and the modest 
[O~III] $\lambda$5007 ratio rule out shocks.  The Balmer
sequence indicates substantial reddening.  After we correct
for reddening, we find that the [O~II] and [S~II] temperatures
agree, placing the nebula between 9,500~K and 12,000~K.
Temperatures like this in low-ionization nebulae
usually signify a hard photoionizing
source that extends into the X-ray band, but the small
He~II $\lambda$4686 / \hbeta ratio we observe shows that the
role of photons $> 54.4$~eV must be very minor.  Hot
stars are the only plausible ionizing sources that remain.

Our finding that hot stars are the most likely ionizing source
agrees with observations of excess blue light and dilution of
stellar absorption features in cooling-flow nebulae.  However,
the hottest O~stars are still too cool to generate the
high temperatures and strong forbidden lines we observe.
In a pure photoionization model, the observed line ratios
require a blackbody-like spectrum exceeding 100,000~K with
a luminosity of a few times $10^{44} \, {\rm erg \, s^{-1}}$.
The photoelectric heating provided by O~stars with effective
temperatures of 50,000~K supplies only about half the 
necessary heating.

Conduction or some other mechanical form of heating might
supplement photoelectric heating in cooling-flow nebulae 
and other LINERs.  At the characteristic ionization parameter
$U \sim 10^{-4}$, the maximum heat flux from the confining
medium $(\sim P v_{\rm th})$ is similar to the photoionizing
energy flux.  Heat sources like this might always be present
in photoionized nebulae, at levels too low to make a difference.
However, in low-ionization nebulae their contributions can grow
significant, boosting the forbidden-line output, especially
if the temperature of the surrounding medium exceeds $10^7$~K,
as in cooling flows or the hot galactic winds expected from
starbursts and AGNs.

\vspace*{2.0em}

M. D. thanks the Carnegie Observatories for the Carnegie Fellowship
that enabled her to gather these data.  The authors would also like 
to acknowldge Daniela Calzetti for donating her reddening-curve 
software, Ari Laor for an enlightening Christmas Eve message,
and Michaela Voit for her continuing inspiration and enthusiasm.

\pagebreak

\pagebreak

\begin{table}
\caption[]{Observed Emission-Line Fluxes\tablenotemark{a} \vspace*{1em}}
\begin{tabular}{lccc}
\hline \bcr
    ~            &   Observed  &   Line Flux \&           &      ~  \\ \bcr
Line ID          &  Wavelength &     1$\sigma$ Error           &   FWHM  \\
          ~            & (\AA) & (10$^{-15}$ erg s$^{-1}$ cm$^{-2}$) & (\AA) \\
\hline \bcr
~ & ~ & ~ & ~ \\ \bcr
OII 3727                 &  4033.65 &       37.   $\pm$ 1.0        & 13.1 \\ \bcr
NeIII 3869	         &  4186.3  &        1.4  $\pm$ 0.10       & 11.5 \\ \bcr
H$\zeta$ 3888.1          &  4208.6  &        0.6  $\pm$ 0.10       & 11.5 \\ \bcr
H$\epsilon$+NeIII 3966   &  4292.1  &        1.95 $\pm$ 0.20       & 13.5 \\ \bcr
SII 4069	         &  4405.0  &        1.5  $\pm$ 0.15       & 13.9 \\ \bcr
H$\delta$ 4101	         &  4437.8  &        1.7  $\pm$ 0.15       & 13.9 \\ \bcr
H$\gamma$ 4340           &  4696.7  &        2.9  $\pm$ 0.3        & 11.3 \\ \bcr
OIII 4363                &    ~     &    $< 0.13 \, (3\sigma)$     &      \\ \bcr
HeII 4686                &  5073.   &        0.26 $\pm$ 0.04       &  ~   \\ \bcr
OIII 4969                &  5367.0  &        1.4  $\pm$ 0.2        & 14.0 \\ \bcr
H$\beta$ 4861            &  5260.4  &        8.4  $\pm$ 0.2        & 14.0 \\ \bcr
OIII 5007                &  5420.8  &        4.8  $\pm$ 0.3        & 14.0 \\ \bcr
NI 5200                  &  5628.5  &        2.7  $\pm$ 0.14       & 14.8 \\ \bcr
OI 6300                  &  6817.8  &       10.9  $\pm$ 0.1        & 14.7 \\ \bcr
OI 6363\tablenotemark{b} &  6886.9  &        2.37 $\pm$ 0.07       & 14.7 \\ \bcr
NII 6548                 &  7086.   &       10.6  $\pm$ 0.09       & 15.5 \\ \bcr
H$\alpha$ 6563           &  7102.35 &       35.4  $\pm$ 0.12       & 15.5 \\ \bcr
NII 6584                 &  7124.48 &       29.5  $\pm$ 0.12       & 15.5 \\ \bcr
HeI 6678                 &  7226.   &	     0.4  $\pm$ 0.1        & 16.2 \\ \bcr
SII 6717                 &  7268.2  &       20.0  $\pm$ 0.1        & 16.2 \\ \bcr
SII 6731                 &  7284.3  &       16.0  $\pm$ 0.11       & 16.2 \\ \bcr
CaII 7290                &  7887.0  &        0.7  $\pm$ 0.1        & 20.0 \\ \bcr
OII+CaII 7320            &  7924.5  &        2.0  $\pm$ 0.2        & 20.0 \\ \bcr

\end{tabular}

\tablenotetext{a}{Uncorrected for interstellar or intrinsic reddening} \bcr
\tablenotetext{b}{Atmospheric absorption ate into the [OI] 6363 feature.}
\end{table}

\pagebreak

\end{document}